%% file: CCS-to-CSPmn_XPRESS-SOS_revised.tex
\documentclass[submission,copyright,creativecommons]{eptcs}
\providecommand{\event}{EXPRESS/SOS 2022}
\usepackage{amssymb}
\usepackage{amsmath}
\usepackage{amsthm}
\usepackage{pifont}
\usepackage{stmaryrd}
\usepackage{xcolor}
\usepackage{cspsymb}
\usepackage{hyperref}
\usepackage{url}
\newtheorem{definition}{Definition}
\newtheorem{theorm}{Theorem}
\newtheorem{example}{Example}
\newtheorem{propty}{Property}

\newcommand{\hideT}{\!\!\hide_{\!_{T}}\!\!}
\newcommand{\hideCSP}{\!\!\hide_{\!_{csp}}\!\!}

\providecommand{\urlalt}[2]{\href{#1}{#2}} 
\providecommand{\doi}[1]{doi:\urlalt{http://dx.doi.org/#1}{#1}}
\title{From CCS to CSP: the m-among-n Synchronisation Approach}%
\author{%
	Gerard Ekembe Ngondi, Vasileios Koutavas, Andrew Butterfield %
	\institute{Trinity College Dublin, Lero - the SFI Software Research Centre}
	\email{gerard.ekembe, vkoutav, andrew.butterfield @tcd.ie}
}%
\def\titlerunning{CCS to CSPmn}
\def\authorrunning{G.Ekembe, V. Koutavas, A. Butterfield}
\begin{document}
\maketitle
\begin{abstract}
We present an alternative translation from CCS to an extension of CSP based on m-among-n synchronisation (called CSPmn). This translation is correct up to strong bisimulation. Unlike the g-star renaming approach (\cite{BEK21}), this translation is not limited by replication (viz., recursion with no nested parallel composition). We show that m-among-n synchronisation can be implemented in CSP based on multiway synchronisation and renaming.
\end{abstract}
\input{Intro}
\input{Backg}
\input{CSPmn}
\input{CCSTransforms}
\input{Gstar2MN}
\input{Conclusion}
	\small 
\paragraph{Acknowledgments.} This work was conducted with the financial support of the Science Foundation Ireland grant 13/RC/2094 and co-funded under the European Regional Development Fund through the Southern and Eastern Regional Operational Programme to Lero - the Irish Software Research Centre (\url{www.lero.ie}). For the purpose of Open Access, the author has applied a CC BY public copyright licence to any Author Accepted Manuscript version arising from this submission.
	\normalsize 
%

%
\appendix
\input{App_ProofCSPmnConserve}
\end{document}

%% file: Intro.tex
\section{Introduction}\label{Intro}
In \cite{BEK21}, the authors present a translation from CCS \cite{ALI05} into
CSP \cite{Schnei00,Ros98}, $ccs2csp$, which is correct up to strong bisimulation
(cf.\ \cite{Glab12}). This means that a CCS process is strong bisimilar to its
CSP translation. $ccs2csp$ has been implemented in Haskell (cf.\
\cite{BEK20git}), which allows using the model-checker FDR \cite{FDR4.2} for analysing translated
CCS terms. In the course of the same work, the authors have proposed an
alternative translation, $ccs2csp_{2}$, correct up to failure equivalence. Both
translations differ in the translation of the prefix term $\tau.P$, translated
into $(tau \then ccs2csp(P))\hideCSP\set{tau}$ in the first case, and
$ccs2csp_{2}(P)$ in the second case.
\par
In this paper we present yet a third alternative, $ccs2csp_{3}$, achieved by first extending CSP with m-among-n synchronisation \cite{Gar99}, from which we can derive multiway (or n-among-n) synchronisation, the default CSP synchronisation mechanism, and binary syncronisation (used in CCS). Then, we translate CCS parallel composition into the binary version of CSP parallel operator. The resulting translation is correct up to strong bisimulation.
\par
The translations in \cite{BEK21} were achieved by hard coding binary synchronisation into CCS before going to CSP. Using a renaming function, $g^{\ast}$, the translations generated unique pairs of indices between any two pairs of complementary prefixes in a parallel composition, e.g., $(a, \bar{a}) \mapsto \set{(a_{12}, \bar{a}_{12}), (a_{13}, \bar{a}_{13})}$. This effectively made synchronising prefix pairs unique. Although these indices were generated in CCS, the $g^{\ast}$-renaming approach shows how to enforce binary synchronisation even in CSP: given a CSP process $P \underset{a}{\parallel} Q \underset{a}{\parallel} R$, to ensure binary synchronisations on $a$, assign unique indices to $a$ accordingly, through renaming. E.g., $P[\set{a_{12}, a_{13}}/a] \parallel Q[a_{12}/a] \parallel R[a_{13}/a]$ ensures that pairs of processes $(P, Q)$ and $(P, R)$ can synchronise respectively, but not $(Q, R)$. This approach, which we call the Gstar approach, has been encoded in the translation tool and the resulting CSP terms can be analysed in FDR immediately.
\par
m-among-n synchronisation \cite{Gar99} demands adding new rules to CSP, hence it would require updating FDR first. In other words, the CSP terms resulting from our new translation, $ccs2csp_{3}$, cannot immediately be analysed in FDR. Nonetheless, function $g^{\ast}$ implements binary synchronisation, hence, can be taken for an implementation of 2-among-n synchronisation.
\par
The Gstar approach does \emph{not} allow translating recursive terms with nested parallelism (or replication). That is because function $g^{\ast}$ needs to generate every synchronisation index so the translation can terminate. With m-among-n synchronisation, we need only one index to separate interleaving from synchronisation, i.e., we map every CCS name unto two CSP events, e.g., $a \mapsto \set{a, a_{S}}$, where $a_{S}$ is the synchronisation event. Therefore, this new translation is not limited by parallel under recursion. \\
\par
Our main contribution in this paper hence is a new translation from CCS into CSP which is correct up to strong bisimulation, is not limited by parallel under recursion, but cannot be immediately analysed with FDR. As a byproduct, we define m-among-n synchronisation for CSP processes. We call the corresponding extension CSPmn. We show that CSPmn preserves CSP axioms by defining m-among-n sysnchronisation in terms of both multiway synchronisation and renaming. The translation from CSPmn into CSP is limited by parallel under recursion as it requires generating unique indices for all possible combinations of synchronising processes.

%% file: Backg.tex
\section{Correct Translation, CCS(Tau), CSP, CCS-to-CSP}
\subsection{Correct Translations}\label{sec:correctTransl}
A correct translation of one language into another is a mapping from the valid
expressions in the first language to those in the second, that preserves their
meaning (for some definition of meaning). Below we recall the two main definitions of correctness from~\cite{Glab12}.

Let $\mathcal{L} = (\mathbb{T}_\mathcal{L}, \llbracket~\rrbracket_\mathcal{L})$ denote a language as a pair of a set $\mathbb{T}_\mathcal{L}$ of valid expressions in $\mathcal{L}$ and a surjective mapping $\llbracket~\rrbracket_\mathcal{L}: \mathbb{T}_\mathcal{L} \then \mathcal{D}_\mathcal{L}$ from $\mathbb{T}_\mathcal{L}$ to some set of meanings $\mathcal{D}_\mathcal{L}$.
Candidate instances of $\llbracket~\rrbracket_\mathcal{L}$ are \emph{traces} and \emph{failures} (cf.\ \cite{Hoa85,Sang12}).

\begin{definition}[Correct Translation up to Semantic Equivalence \cite{Glab12}]\label{def:correct-upto}
A translation $\mathsf{T}: \mathbb{T}_\mathcal{L} \then \mathbb{T}_\mathcal{L'}$ is correct up to a semantic equivalence $\approx$ on $\mathcal{D}_\mathcal{L} \cup \mathcal{D}_\mathcal{L'}$ when 
$\llbracket E \rrbracket_\mathcal{L} \approx
	\llbracket\mathsf{T}(E)\rrbracket_\mathcal{L'}$ for all $E \in \mathbb{T}_\mathcal{L}$.
\end{definition}
\par
Operational correspondence allows matching the transitions of two processes,
which can help determine the appropriate relation (semantic equivalence) between
a term and its translation. Let the operational semantics of $\mathcal{L}$ be
defined by the labelled transition system $(\mathbb{T}_\mathcal{L},
Act_\mathcal{L}, \xrightarrow{~~}_\mathcal{L})$, where $Act_\mathcal{L}$ is the
set of labels and $E\xrightarrow{\lambda}_\mathcal{L} E'$ defines transitions
with $E,E'\in\mathbb{T}_{\mathcal{L}}$ and $\lambda\in Act_{\mathcal{L}}$.

\begin{definition}[Labelled Operational Correspondence, \cite{Fu10,Pet19}]\label{def:opCorrespond}
Let $\mathsf{T}: \mathbb{T}_\mathcal{L} \then \mathbb{T}_\mathcal{L'}$ be a mapping from the expressions of a language $\mathcal{L}$ to those of a language $\mathcal{L'}$, and let $\mathsf{f}: Act_\mathcal{L} \then Act_\mathcal{L'}$ be a mapping from the labels of $\mathcal{L}$ to those of $\mathcal{L'}$. A translation $\langle \mathsf{T}, \mathsf{f} \rangle$ is operationally corresponding w.r.t.\ a semantic equivalence $\approx$ on $\mathcal{D}_\mathcal{L} \cup \mathcal{D}_\mathcal{L'}$ if it is:
	\begin{itemize}
	\item Sound: $\forall E, E': E \xrightarrow{\lambda}_\mathcal{L} E'$ imply that $\exists F: \mathsf{T}(E) \xrightarrow{\mathsf{f}(\lambda)}_\mathcal{L'} F$ and $F \approx \mathsf{T}(E')$
	\item Complete: $\forall E, F: \mathsf{T}(E) \xrightarrow{\lambda'}_\mathcal{L'} F$ imply that $\exists E': E \xrightarrow{\lambda}_\mathcal{L} E'$ and $F \approx \mathsf{T}(E') \land \lambda' = \mathsf{f}(\lambda)$
	\end{itemize}
\end{definition}
The previous two definitions coincide when the semantic equivalence $\approx$ is strong bisimulation (Def.\ref{def:strongbisim}) and $\mathsf{f}$ is the identity.
\subsection{CCS, CCSTau}
\begin{table}[t]
\caption{SOS rules for CCS}\label{tab:CCSrules}%
\begin{align*}
Prefix:~~& \alpha{.P} \xrightarrow{\alpha} P & %
Sum:~~& \frac{P \xrightarrow{\alpha} P'}{P+Q \xrightarrow{\alpha} P'} &
Par:~~& \frac{P \xrightarrow{\alpha} P'}{P|Q \xrightarrow{\alpha} P'|Q} 
\\%
Com:~~& \frac{P \xrightarrow{\overline{a}} P' \quad Q \xrightarrow{a} Q'}{P|Q \xrightarrow{\tau} P'|Q'} &
Res:~~& \frac{P \xrightarrow{\alpha} P' \quad \alpha \notin B}{P\restriction{B} \xrightarrow{\alpha} P'\restriction{B}} &
Rec:~~& \frac{P[\mu{X}.P / X] \xrightarrow{\alpha} P'}{\mu{X}.P \xrightarrow{\alpha} P'} \\%
\end{align*}
\end{table}
\paragraph{CCS.} CCS (Calculus of Communicating Systems) \cite{Mil89,ALI05} is a process algebra that allows reasoning about concurrent systems. CCS represents programs as \emph{processes}, whose behaviour is determined by rules specifying their possible execution steps. The syntax of CCS processes is defined by the following BNF:
\begin{align*} 
CCS &::= 0 ~|~ \alpha.P ~|~ P + Q ~|~ P | Q ~|~ P\restriction{B} ~|~ \mu{X}.P \\
\alpha &::= \tau ~|~ \overline{a} ~|~ a 
\end{align*}
\par 
Let $\mathcal{N}$ denote an infinite set of \emph{names}; 
let $a, b, c, ...$ range over $\mathcal{N}$. 
Let $\overline{\mathcal{N}} = \{\bar{a} | a \in \mathcal{N}\}$ denote the set of conames. 
Let $\overline{\overline{a}} = a$. Let $\mathcal{L} = \mathcal{N} \cup \overline{\mathcal{N}}$ denote the set of all possible labels. The set of labels of a process $P$ is denoted by $\mathcal{L}(P)$ (\cite[Def.2, p52]{Mil89}). 
Let $\tau$ denote the silent or invisible action. 
Let $Act = \mathcal{N} \cup \overline{\mathcal{N}} \cup \{\tau\}$ denote the set of all possible actions that a process can perform. Let $\alpha, \beta, ..$ range over $Act$. 
The SOS semantics of CCS are given in Table \ref{tab:CCSrules}.
\par 
Informally: 
$0$ (or $NIL$) is the process that performs no action. 
$\alpha{.P}$ is the process that performs an action $\alpha$ and then behaves like $P$. 
$P + Q$ is the process that behaves either like $P$ or like $Q$. 
$P|Q$ is the process that executes $P$ and $Q$ in parallel: if both $P$ and $Q$ can engage in an action $a$ then, their execution corresponds to interleaving, e.g.\ $a.0 | a.0 \equiv a.a.0$; if $P$ can engage in action $a$, $Q$ in the complementary action $\bar{a}$, then, either $P$ and $Q$ interleave on $a$ or they synchronise and the result of synchronisation is the invisible action $\tau$, e.g.\ $a.0 | \bar{a}.0 \equiv a.\bar{a}.0 + \bar{a}.a.0 + \tau.0$. 
$P\restriction{B}$ is the process that cannot engage in actions in $B$ except for synchronisation, e.g., $(a.0 | \bar{a}.0)\restriction\{a\} \equiv \tau.0$, $(a.0)\restriction\{a\} \equiv 0$. 
$\mu{X}.P$ is the process that executes $P$ recursively. \\
\par 
Equivalence based on bisimulations is the preferred choice for discriminating among CCS processes. We will use strong bisimulation to prove the correctness of our translation. 
\begin{definition}[Strong Bisimulation \cite{Sang12, Mil89}]\label{def:strongbisim}
A strong bisimulation is a symmetric binary relation $\mathcal{R}$ on processes satisfying the following: $P \mathcal{R} Q$ and $P \xrightarrow{\alpha} P'$ imply that
\[ \exists Q': Q \xrightarrow{\alpha} Q' \land P' \mathcal{R} Q' \]
$P$ is strong bisimilar to $Q$, written $P \sim Q$, if $P \mathcal{R} Q$ for some strong bisimulation $\mathcal{R}$.
\end{definition}
\paragraph{CCSTau.} CCSTau \cite{BEK21} extends CCS with visible synchronisations, viz., the result of synchronisation on a pair $(a, \bar{a})$ is the visible action $\tau[a,\bar{a}]$ instead of the visible action $\tau$. This makes it easier to guarantee that when two processes synchronise in CCS(Tau), their CSP translation also synchronises. 
The syntax of CCSTau processes is defined by the following grammar:
\begin{align*} 
P, Q, R &::= 0 ~~|~~ \alpha.P ~~|~~ P + Q ~~|~~ P |_{_{T}} Q ~~|~~ P\restriction{B} ~~|~~ \rec{X}{P} ~~|~~ P\hideT{B} ~~|~~ X \\
\alpha &::= \tau ~~|~~ \overline{a} ~~|~~ a \\
\beta &::= \alpha ~~|~~  \tau[a|\overline{a}]
\end{align*}
The parallel operator in CCSTau is denoted $|_{_{T}}$. CCSTau also defines a hiding operator, denoted ~$\hideT$~, which can hide all actions including $\tau[a,\bar{a}]$ actions. The restriction operator behaves as in CCS, does not apply to $\tau[a,\bar{a}]$ actions.
Rules for these operators are given hereafter:
\begin{align*}
Par:&~ \dfrac{P \xrightarrow{\beta} P'}{P |_{_{T}} Q \xrightarrow{\beta} P' |_{_{T}} Q} 
&\quad
Com:~~ \dfrac{P \xrightarrow{\overline{a}} P' \quad Q \xrightarrow{a} Q'}
	{P |_{_{T}} Q \xrightarrow{\tau[\overline{a}|a]} P' |_{_{T}} Q'} \\
Res:&~~ \dfrac{P \xrightarrow{\beta} P' \quad \beta = \tau[\overline{a}|a] \text{ or } \beta \notin B}
	{P\restriction{B} \xrightarrow{\beta} P'\restriction{B}} \\%
Hide:&~~ \dfrac{P \xrightarrow{\beta} P' \quad \beta \notin B}{P\hideT{B} \xrightarrow{\beta} P'\hideT{B}}
	&\quad \dfrac{P \xrightarrow{\beta} P' \quad \beta \in B}{P\hideT{B} \xrightarrow{\tau} P'\hideT{B}}
\end{align*}
All other CCS operators are also CCSTau operators. 
\paragraph{CCS-to-CCSTau.} Translation function $c2ccs\tau$ \cite{BEK21} translates CCS processes into CCSTau, is correct up to strong bisimulation. For any CCS process $P$ other than CCS-parallel operator, $c2ccs\tau(P) = P$. For the parallel operator: 
	\footnote{The set of labels of a CCS process $P$, $\mathcal{L}(P)$, corresponds to the set of events $\mathcal{A}(Q)$ for a CSP process $Q$. }
\begin{align*}
c2ccs\tau(P | Q) ~\defs~ (c2ccs\tau(P) &|_{_{T}} c2ccs\tau(Q))\hideT\{\tau[a|\overline{a}] ~|~ a \in \mathcal{L}(P), \bar{a} \in \mathcal{L}(Q)\} \tag{c2ccs$\tau$-par-def}\label{c2ccstau-par-def}
\end{align*}
\subsection{CSP}
\begin{table}[t]
\caption{SOS rules for CSP \cite{Schnei00} }\label{tab:CSPrules} %
\begin{align*}
Prefix:~~& (a \leadsto P) \xrightarrow{a} P 
	&Skip:~~ SKIP \xrightarrow{\tick} STOP \\
IntChoice:~~& P_{1} \intchoice P_{2} \xrightarrow{\tau} P_{1} 
	&P_{1} \intchoice P_{2} \xrightarrow{\tau} P_{2} \\
ExtChoice:~~& \frac{P_{1} \xrightarrow{a} P'}{P_{1} \extchoice P_{2} \xrightarrow{a} P'} 
	&\frac{P_{1} \xrightarrow{\tau} P'}{P_{1} \extchoice P_{2} \xrightarrow{\tau} P' \extchoice P_{2}} \\[+0.85ex]
IfacePar:~~& \frac{P_{1} \xrightarrow{a} P' \quad [a \notin B^{\tick} ] }{P_{1} \underset{B}{\parallel} P_{2} \xrightarrow{a} P' \underset{B}{\parallel} P_{2}} 
	&\frac{P_{1} \xrightarrow{a} P'_{1} \quad P_{2} \xrightarrow{a} P'_{2} \quad [a \in B^{\tick}]}{P_{1} \underset{B}{\parallel} P_{2} \xrightarrow{a} P'_{1} \underset{B}{\parallel} P'_{2}} \\
Hide:~~& \frac{P \xrightarrow{a} P' \quad [a \notin B]}{P\hide{B} \xrightarrow{a} P'\hide{B}} 
	&\frac{P \xrightarrow{a} P' \quad [a \in B]}{P\hide{B} \xrightarrow{\tau} P'\hide{B}} \\
FwdRen:~~& \frac{P \xrightarrow{a} P'}{f(P) \xrightarrow{f(a)} f(P')} 
&
\frac{P \xrightarrow{\tau} P'}{f(P) \xrightarrow{\tau} f(P')} \\
Rec:~~& \frac{P \xrightarrow{\mu} P' \quad [N = P] }{N \xrightarrow{\mu} P'} 
\end{align*}
\end{table}
CSP (Communicating Sequential Processes) \cite{Hoa85, Schnei00} is a process algebra that allows reasoning about concurrent systems. In CSP, a (concurrent) program is represented as a \emph{process}, whose behaviour is entirely determined by the possible actions of the program, represented as \emph{events}. The set of events that a process $P$ can possibly perform is denoted by $\mathcal{A}(P)$. Event $\tau$ denotes invisible actions, hidden from the environment; event $\tick$ denotes successful termination, by opposition say to deadlock and abortion. Both denotational and operational semantics have been defined for CSP processes, in terms of traces. The syntax of some CSP processes is defined by the following BNF:
\begin{align*}
CSP &::= SKIP ~|~ STOP ~|~ \alpha \leadsto P ~|~ P \intchoice Q ~|~ P \extchoice Q ~|~ P \underset{B}{\parallel} Q ~|~
	f(P) ~|~ P\hide{B} ~|~ \mu{X}.P \\
\alpha &::= a ~|~ a?x ~|~ a!m
\end{align*}
\par
The SOS semantics of CSP processes are given in Table \ref{tab:CSPrules}. Informally: $SKIP$  is the process that refuses to engage in any event, terminates immediately, and does not diverge. 
$STOP$ is the process that is unable to interact with its environment. 
$\alpha \leadsto P$ is the process that first engages in event $\alpha$ then behaves like $P$.
$P \extchoice Q$ is the process that behaves like $P$ or $Q$, where the choice is decided by the environment.
$P \underset{B}{\parallel} Q$ behaves like the parallel execution of $P$ and $Q$ where the latter must both synchronise on the set of events $B$. When $B = \{\}$, we say that $P$ and $Q$ interleave, denoted by $P \interleave Q$; if $B = \mathcal{A}(P) \cap \mathcal{A}(Q)$ we also write $P \parallel Q$. 
$f(P)$ engages in $f(a)$ whenever $P$ engages in $a$. 
$P\hide{B}$ is the process that engages in all events of $P$ except those in $B$. 
$\mu{X}.P$ is the process that executes $P$ recursively. \\
\par 
Equivalence based on (enriched versions of) traces is the preferred choice for distinguishing CSP processes. We kindly refer the reader to \cite{Hoa85,Schnei00} for details. 
\subsection{CCS-to-CSP Translation}
\paragraph*{Notation.} Given two functions, say $f_{1}$ and $f_{2}$, $f_{1} \circ f_{2}$ denotes functional composition, viz., $f_{1}(f_{2})$. \\

In this section, we present $ccs2csp$ {\cite{BEK21}}, the translation from CCS-to-CSP, correct up to strong bisimulation. 
\begin{definition}[$ccs2csp$ {\cite{BEK21}}]\label{def:ccs2csp}
Let $P$ be a CCS process. Then:
\begin{align*}
ccs2csp(P) ~&\defs~ ai2a \circ (t2csp \circ c2ccs\tau (P))\hideCSP\set{a_{ij} | a_{ij} \in \mathcal{A}\big(t2csp(c2ccs\tau(P))\big)} \\
t2csp(P) ~&\defs~ (tl \circ conm \circ g^{\ast}_{\nullset} \circ ix(P))\hideCSP\set{tau} \\
g^{\ast}_{S} ~&\defs~ \set{\tau \mapsto \tau, a_{i} \mapsto \set{a_{i}} \cup \set{a_{ij}| \bar{a}_{j} \in S, i < j} \cup \set{a_{ji}| \bar{a}_{j} \in S, j < i} } \\
conm ~&\defs~ \{\tau \mapsto \tau, a_{i} \mapsto a_{i}, \bar{a}_{i} \mapsto \bar{a}_{i}, a_{ij} \mapsto a_{ij},\bar{a}_{ij} \mapsto a_{ij} \} \\
ai2a ~&\defs~ \{a_{i} \mapsto a\}
\end{align*}
\end{definition}
where $ix$ generates unique indexed prefixes such that a name $b$ maps to a set of indexed names $b_{i}, i \geq 1$; $g^{\ast}$ generates unique double-indexed names for every pair of synchronising names; $conm$ renames every synchronising coname into the corresponding name (so they can synchronise in CSP); and $tl$ translates CCS operators into corresponding CSP operators. We kindly refer the reader to \cite{BEK21} for details.
\begin{example}[{\cite{BEK21}}]\label{eg:ccs2csp-par}
The translation of CCS binary synchronisation into CSP can be illustrated succinctly as follows:
\begin{align*}
& ccs2csp(a.0 | \bar{a}.0) \tag{ccs2csp-def}\\
=& 
ai2a \circ t2csp\big(c2ccs\tau(a.0 | \bar{a}.0)\big)\hideCSP\set{a_{ij} | ..} \tag{c2ccs$\tau$-par-def}\\
=& 
ai2a \circ t2csp\big((a.0 |_{_{T}} \bar{a}.0)\hideT\{\tau[a|\bar{a}]\}\big)\hideCSP\set{a_{ij} | ..} \tag{t2csp-def}\\
=& 
ai2a \circ tl \circ conm \circ g^{\ast}(\nullset, ix\big((a.0 |_{_{T}} \bar{a}.0)\hideT\{\tau[a|\bar{a}]\}\big))\hideCSP\set{tau}\hideCSP\set{a_{ij} | ..} \tag{ix-def}\\
=& 
ai2a \circ tl \circ conm \circ g^{\ast}\big((a_{1}.0 |_{_{T}} \bar{a}_{2}.0)\big)\hideCSP\set{tau}\hideCSP\set{a_{ij} | ..} \tag{gstar-def}\\
=& 
ai2a \circ tl \circ conm \big((a_{1}.0 + a_{12}.0) |_{_{T}} (\bar{a}_{2}.0 + \bar{a}_{12}.0)\big)\hideCSP\set{tau}\hideCSP\set{a_{12}} \tag{conm-def}\\
=& 
ai2a \circ tl \big((a_{1}.0 + a_{12}.0) |_{_{T}} (\bar{a}_{2}.0 + a_{12}.0)\big)\hideCSP\set{tau, a_{12}} \tag{tl-def}\\
=& 
ai2a \circ \big((a_{1} \extchoice a_{12} \leadsto STOP) \underset{\{a_{12}\}}{\parallel} (\bar{a}_{2} \extchoice a_{12} \leadsto STOP)\big)\hideCSP\set{tau, a_{12}}  \tag{ai2a-def}\\
=& 
\big((a \extchoice a_{12} \leadsto STOP) \underset{\{a_{12}\}}{\parallel} (\bar{a} \extchoice a_{12} \leadsto STOP)\big)\hideCSP\set{tau, a_{12}}  %
\end{align*}
\end{example}
In CCS, a name can be used both for interleaving and for synchronisation. This is reflected in the translation above by generating indexed names $a_{1}$ and $\bar{a}_{2}$ for interleaving; then for the synchronisation pair $(a_{1}, \bar{a}_{2})$, a unique synchronisation name $a_{12}$ is generated. More generally, there will be as many $a_{ij}$ synchronisation names as there are of synchronisation on name $a$. 
\par 
In the next section, we extend CSP with m-among-n synchronisation, then derive 2-among-n (binary) synchronisation. In the end, we will be able to translate CCS binary synchronisation into CSP binary synchronisation.

%% file: CSPmn.tex
\section{CSP plus m-among-n Synchronisation}\label{sec:cspmn}
Multiway synchronisation in CSP is \emph{maximal}, viz., all processes that can synchronise \emph{must} synchronise. This is also called the \emph{maximal (or n-ary) coordination} paradigm (\cite{Gar99}): if $n$ processes are ready to synchronise on event $a$, then all $n$ processes must synchronise together. Can we generalise this to allow only m-among-n ($2 \leq m \leq n$) processes to synchronise instead? If the answer is yes then binary synchronisation can be defined as 2-among-n coordination and n-ary synchronisation as n-among-n coordination. Garavel and Sighireanu \cite{Gar99} define $m/n$ coordination for the language E-LOTOS. \\
\par 
First, let us generalise CSP (n-ary) interface parallel operator (\cite{Schnei00}).
	
\begin{align*}
IndxIfacePar:~~& \frac{P_{j} \xrightarrow{a} P' \quad [a \notin B^{\tick}, k \neq j ] }
	{\underset{B}{\Parallel} P_{i} \xrightarrow{a} (\underset{B}{\Parallel}P_{k}) \underset{B}{\parallel}P'} %
	&%
\frac{P_{1} \xrightarrow{a} P'_{1} \quad ... \quad P_{n} \xrightarrow{a} P'_{n} \quad [a \in B^{\tick}]}
	{\underset{B}{\Parallel} P_{i} \xrightarrow{a} \underset{B}{\Parallel}P'_{i}}
\end{align*}
\begin{definition}[$a\#{m}$ clause \cite{Gar99}]\label{def:axm-clause}
Let $I = \{1, .., n\}, n \in \nats, n \geq 2$. Let $m$ be a natural number  in the range $2, .., n$ associated to an $a$-event such that a clause $a\#{m}$ denotes that $m$ processes are allowed to synchronise on event $a$ at once. Each clause $\#m$ is optional: if omitted, $m$ has default value $n$. 
\end{definition}
The rules for $m/n$ indexed interface paralell composition are given hereafter.\footnote{The rules in \cite{Gar99} use a different rule format than CSP rules: they use predicates.} %
\begin{align*}
M{\!}/{\!}N{\!\!}-{\!\!}IndxIfacePar:~~&\frac{P_{j} \xrightarrow{a} P' \quad [a\#m \notin B^{\tick} \times \{2, .., n\}, k \neq j]}
	{\underset{B \times \{2, .., n\}}{\Parallel} P_{i} \xrightarrow{a} (\underset{B \times \{2, .., n\}}{\Parallel}P_{k}) \underset{B \times \{2,..,n\}}{\parallel}P'} \\[+0.5ex]
	&%
\frac{P_{1} \xrightarrow{a} P'_{1}~~ ... ~~P_{n} \xrightarrow{a} P'_{n} \quad [a\#m \in B^{\tick} \times \{2, .., n\}, j \in J, k \neq j]}
	{\underset{B \times \{2, .., n\}}{\Parallel} P_{i} \xrightarrow{a} \underset{\{J \subseteq I | card(J) = m\}}{\Intchoice} \bigg((\underset{B \times \{2, .., n\}}{\Parallel}P_{k}) \underset{B \times \{2,..,n\}}{\parallel} (\underset{B \times \{2, .., n\}}{\Parallel}P'_{j}) \bigg)} 
\end{align*} 
We can then derive binary-only synchronisation by imposing that \emph{every} event in set $B$ allows 2(only)-among-n processes to synchronise.
\begin{align*}
2{\!}/{\!}N{\!\!}-{\!\!}IndxIfacePar:~~
	&%
\frac{P_{1} \xrightarrow{a} P'_{1}~~ ... ~~P_{n} \xrightarrow{a} P'_{n} \quad [a\#2 \in A^{\tick} \times \{2\}, j \in J, k \neq j]}
	{\underset{B \times\{2\}}{\Parallel} P_{i} \xrightarrow{a} \underset{\{J \subseteq I | card(J) = 2\}}{\Intchoice} \bigg((\underset{B \times\{2\}}{\Parallel}P_{k}) \underset{B \times \{2\}}{\parallel} (\underset{B \times \{2\}}{\Parallel}P'_{j}) \bigg)} 
\end{align*}
Similarly, we derive n-ary-only synchronisation by imposing that \emph{every} event in set $B$ allows n-among-n processes to synchronise. We easily verify that rules N/N-IndxIfacePar and IndxIfacePar (synchronisation) are the same. %
\begin{align*}
N{\!}/{\!}N{\!\!}-{\!\!}IndxIfacePar:~~ &\frac{P_{1} \xrightarrow{a} P'_{1}~~ ... ~~P_{n} \xrightarrow{a} P'_{n} \quad [a\#n \in B^{\tick} \times \{n\}]}
	{\underset{B \times \{n\}}{\Parallel} P_{i} \xrightarrow{a} \underset{B \times \{n\}}{\Parallel}P'_{i}} 
\end{align*}
\paragraph{Correctness of M/N-IndxIfacePar rule.} Let us call CSPmn the extension of CSP with m-among-n synchronisation. We argue here that CSPmn is a conservative extension of CSP, i.e., CSPmn preserves the axioms of CSP. 
\par 
The proof method is suggested to us by function $g^{\ast}$ \cite{BEK21}. For binary synchronisation, select process pairs that must synchronise and assign them a unique synchronisation index. 
E.g.,
$$ a \underset{a\#2}{\parallel} a \underset{a\#2}{\parallel} a
\quad\text{maps to}\quad 
(a_{12} \extchoice a_{13}) \parallel (a_{12} \extchoice a_{23}) \parallel (a_{13} \extchoice a_{23})
$$
Then, for $m$ processes to synchronise among $n$, generate a unique index for all possible combinations of $m$ processes among $n$, e.g.,
\begin{align*}
a \underset{a\#2}{\parallel} a \underset{a\#2}{\parallel} a \underset{a\#2}{\parallel} a 
\quad&\text{maps to}\quad
(a_{12} \extchoice a_{13} \extchoice a_{14}) \parallel (a_{12} \extchoice a_{23} \extchoice a_{24}) \parallel (a_{13} \extchoice a_{23} \extchoice a_{34}) \parallel \\
	&\hspace{3cm} (a_{14} \extchoice a_{24} \extchoice a_{34})
\\
a \underset{a\#3}{\parallel} a \underset{a\#3}{\parallel} a \underset{a\#3}{\parallel} a 
\quad&\text{maps to}\quad
(a_{123} \extchoice a_{124} \extchoice a_{134}) \parallel (a_{123} \extchoice a_{124} \extchoice a_{234}) \parallel (a_{123} \extchoice a_{134} \extchoice a_{234}) \parallel \\
	&\hspace{3cm} (a_{124} \extchoice a_{134} \extchoice a_{234}) 
\\
a \underset{a\#4}{\parallel} a \underset{a\#4}{\parallel} a \underset{a\#4}{\parallel} a 
\quad&\text{maps to}\quad 
a_{1234} \parallel a_{1234} \parallel a_{1234} \parallel a_{1234}
\end{align*}


From what precedes, there exists a relational renaming, say $G$, such that
$$
\underset{a\#m, j}{\Parallel}P_{j} \sim \underset{G(a), j}{\Parallel} P_{j}[G(a)/a]
$$
We can thus define (CSPmn parallel operator) $\underset{a\#m}{\parallel}$ in terms of both (CSP parallel operator) $\underset{a}{\parallel}$ and (CSP relational renaming) $G(a)$. Therefore, CSPmn is a conservative extension of CSP, viz., preserves CSP axioms (cf. Appendix \ref{app:proof-cspmn-corr} for a full proof).

%% file: CCSTransforms.tex
\section{CCSTau Transformations}\label{sec:ccs-transforms}
\begin{figure}[h]\center
$\boxed{CCS} \xrightarrow{c2ccs\tau}\boxed{CCSTau} \xrightarrow{g^{2}}
	\xrightarrow{conm}\boxed{CCSTau}\xrightarrow{tl_{3}}
	\xrightarrow{\hide\set{tau}} \xrightarrow{\hide\set{a_{S}}} \boxed{CSPmn}$
	
	\caption{CCS-to-CSPmn Translation workflow}\label{fig:transl-workflow}
\end{figure}
The different stages of our translation are shown in Fig.~\ref{fig:transl-workflow}.
\paragraph{Pairwise vs.\ Multiway Synchronisation}
Recall, a CCSTau name has both interleaving and synchronisation semantics. We hence have to generate two distinct CSP events for a single CCS name. 
Also, it is possible to hide $\tau[a_{}|\bar{a}_{}]$ synchronisation actions in CCSTau (typically, to obtain a CCS process---cf.\ Def.\ref{c2ccstau-par-def}). Then, it will be convenient to ignore them. 
 Let $g^{2}$ define the function that generates a synchronisation name for any CCS name. 
\begin{definition}[$g^{2}(\alpha)$]\label{def:g(alpha):m/n}
\begin{align*}
g^{2}(S, \tau) &\defs~~ \tau 
	&\quad g^{2}(S, a) &\defs~~ \set{a} \cup \set{a_{S} ~|~ \bar{a} \in S} \\
g^{2}(S, \tau[a | \bar{a}]) &\defs~~ \set{\tau[a,\bar{a}]}
	&\quad g^{2}(S, B) &\defs~~ \set{g^{2}(S, a) ~|~ a \in B, \bar{a} \in S}
\end{align*}
\end{definition}
\par 
Given a set of names generated by $g^{2}$, $a$-names denote interleaving, whilst $a_{S}$-names denote synchronisation. 
The application of $g^{2}$ to processes is given hereafter.
\begin{definition}[$g^{2}(P)$]\label{def:g(proc)}
Let $P$ be a CCS process. Let $g^{2}(P) \defs g^{2}(\nullset, P)$.
\begin{align*} 
&\begin{aligned} 
g^{2}(S, 0) &\defs~ 0 \\
g^{2}(S, \alpha.P) &\defs~ \underset{b \in g^{2}(S, \alpha)}{\Sigma} b.g^{2}(S, P) \\
g^{2}(S, P + Q) &\defs~ g^{2}(S, P) + g^{2}(S, Q) \\
g^{2}(S, P |_{_{T}} Q)  &\defs~ g^{2}(S \cup \mathcal{A}(Q), P) |_{_{T}} g^{2}(S \cup \mathcal{A}(P), Q) 
\end{aligned} 
&\begin{aligned}
g^{2}(S, P\restriction{B}) &\defs~ g^{2}(S, P) \restriction g^{2}(S, B) \\
g^{2}(S, P\hideT{B}) &\defs~ g^{2}(S, P)\hideT g^{2}(S \cup B, B) \\
g^{2}(S, \rec{X}{P}) &\defs~ \rec{X}{g^{2}(S, P)} \\%
g^{2}(S, X) &\defs~ X 
\end{aligned} 
\end{align*}
\end{definition}
Note the difference between restriction and hiding. Names $g^{2}(S, B)$ are generated between a process and its environment. Only those names will be restricted, understood that (restricted) $B$ names cannot interact with their environment. Internal synchronisation on $B$ names, however, will not be restricted (until later in CSP). In contrast, for hiding, internal synchronisation on $B$ must be hidden as well, hence we hide names $g^{2}(S \cup B, B)$ instead.
\begin{example}\label{eg:gstar-1}
Let us illustrate the translation of restriction.
\begin{align*}
& g^{2}\big(\nullset, (a.0 |_{_{T}} \bar{a}.0)\restriction\{a\}\big) \tag{g2-def}\\
=&~ 
g^{2}\big(\nullset, a_{}.0 |_{_{T}} \bar{a}_{}.0\big)\restriction g^{2}(\{\}, \{a\}) \tag{g2-res-def}\\
=&~ 
\big(g^{2}(\{\bar{a}_{}\}, a_{}.0) |_{_{T}} g^{2}(\{a_{}\}, \bar{a}_{}.0)\big)\restriction\set{a} \tag{g2-par-def}\\
=&~ 
\big((a_{}.0 + a_{S}.0) |_{_{T}} (\bar{a}_{}.0 + \bar{a}_{S}.0)\big)\restriction \{a\} %
\end{align*}
Contrast with hiding, which hides both $a$ and $a_{S}$. (Recall ~ $\hideT\set{a} = \hideT\set{a, \bar{a}}$.)
\begin{align*}
& g^{2}\big(\nullset, (a.0 |_{_{T}} \bar{a}.0)\hideT\{a\}\big) \tag{hide-def}\\
=&~ 
g^{2}\big(\nullset, (a.0 |_{_{T}} \bar{a}.0)\hideT\{a,\bar{a}\}\big) \tag{g2-def}\\
=&~ 
g^{2}\big(\nullset, a_{}.0 |_{_{T}} \bar{a}_{}.0\big)\hideT g^{2}(\{a,\bar{a}\}, \{a,\bar{a}\}) \tag{g2-hide-def}\\
=&~ 
\big(g^{2}(\{\bar{a}_{}\}, a_{}.0) |_{_{T}} g^{2}(\{a_{}\}, \bar{a}_{}.0)\big)\hideT\set{a,\bar{a}, a_{S}, \bar{a}_{S}} \tag{g2-par-def, hide-def}\\
=&~ 
\big((a_{}.0 + a_{S}.0) |_{_{T}} (\bar{a}_{}.0 + \bar{a}_{S}.0)\big)\hideT \{a,a_{S}\} %
\end{align*}
Finally, consider hiding the synchronisation action $\tau[a|\bar{a}]$, this turns out to be vacuous.
\begin{align*}
& g^{2}\big(\nullset, (a.0 |_{_{T}} \bar{a}.0)\hideT\{\tau[a|\bar{a}]\}\big) \tag{g2-def}\\
=&~ g^{2}\big(\nullset, a_{}.0 |_{_{T}} \bar{a}_{}.0\big)\hideT g^{2}(\{a\}, \{\tau[a|\bar{a}]\}) \tag{g2-hide-def}\\
=&~ \big(g^{2}(\{\bar{a}_{}\}, a_{}.0) |_{_{T}} g^{2}(\{a_{}\}, \bar{a}_{}.0)\big)\hideT\set{\tau[a|\bar{a}]} \tag{g2-par-def}\\
=&~ \big((a_{}.0 + a_{S}.0) |_{_{T}} (\bar{a}_{}.0 + \bar{a}_{S}.0)\big)\hideT \{\tau[a|\bar{a}]\} %
\end{align*}
\end{example}
\paragraph{Parallel Composition.} In CSP, synchronisation pairs $(a_{S}, \bar{a}_{S})$ will not be able to synchronise. We hence update the coname function to translate conames into names.
\begin{definition}[$conm$]\label{def:conm}
$ conm ~\defs~ \{\tau \mapsto \tau, a \mapsto a, \bar{a} \mapsto \bar{a}, a_{S} \mapsto a_{S}, \bar{a}_{S} \mapsto a_{S} \} $.
\end{definition}
\paragraph{Link CCSTau-to-CSPmn}\label{ss:link-ccstau-csp:m/n}
In \cite{BEK21}, function $tl$ translates CCSTau operators into CSP operators, without consideration for differences in their respective alphabets. Hereafter, we define $tl_{3}$, to map CCS binary synchronisation into CSPmn binary synchronisation. All other operators are translated as before, viz., $tl_{3}(P) = tl(P)$ for all process expressions other than parallel composition. Additionally, because of the possibility to hide $\tau[a,\bar{a}]$ synchronisation actions in CCSTau, we translate CCSTau hiding operator also, translation which was not needed for $tl$.
\begin{definition}[$tl_{3}$]\label{def:tl3}
Let $tau$ be a CSP event that cannot synchronise.
\begin{align*}
&\begin{aligned}
tl_{3}(0) ~&\defs~ STOP \\
tl_{3}(\tau.P) ~&\defs~ tau \leadsto tl_{3}(P) \\
tl_{3}(a.P) ~&\defs~ a \leadsto tl_{3}(P) \\
tl_{3}(P\restriction{B}) ~&\defs~ tl_{3}(P)\restriction_{csp}{B} \\
tl_{3}(P\hideT{B}) ~&\defs~ tl_{3}(P)\hideCSP{B} \\
\end{aligned}
&\begin{aligned}
tl_{3}(P + Q) ~&\defs~ tl_{3}(P) \extchoice tl_{3}(Q) \\
tl_{3}(P |_{_{T}} Q) ~&\defs~~ tl_{3}(P) \underset{\{a\#2 | a \in \mathcal{A}(P) ~\cap~ \mathcal{A}(Q) \} }{\parallel} tl_{3}(Q) \\
tl_{3}(\rec{X}{P}) ~&\defs~ \rec{X}{tl_{3}(P)} \\
tl_{3}(X) ~&\defs~ X
\end{aligned}
\end{align*}
\end{definition}
Note that $tl_{3}(P\hideT\set{\tau[a|\bar{a}]}) = tl_{3}(P)\hideCSP\set{\tau[a|\bar{a}]} = tl_{3}(P)$, since $\tau[a|\bar{a}]$ actions do not occur in the translated term, $tl_{3}(P)$. This is necessary, as illustrated subsequently.
\begin{example}
CCS process $a ~|~ \bar{a} ~|~ a$, by $c2ccs\tau$, corresponds to CCSTau process 
$$\big((a |_{_{T}} \bar{a})\hideT\set{\tau[a|\bar{a}]} |_{_{T}} a \big)\hideT\set{\tau[a,\bar{a}]}$$
By $g^{2}$, this becomes process
$$\big(((a + a_{S}) |_{_{T}} (\bar{a} + \bar{a}_{S}))\hideT\set{\tau[a|\bar{a}]} |_{_{T}} (a + a_{S}) \big)\hideT\set{\tau[a|\bar{a}]}$$ 
Then, by $tl_{3}$, it becomes 
\begin{align*}
&\big(((a \extchoice a_{S}) \underset{a_{S}\#2}{\parallel} (\bar{a} \extchoice \bar{a}_{S}))\hideCSP\set{\tau[a|\bar{a}]} \underset{a_{S}\#2}{\parallel} (a \extchoice a_{S}) \big)\hideCSP\set{\tau[a|\bar{a}]} \\
=&~ (a \extchoice a_{S}) \underset{a_{S}\#2}{\parallel} (\bar{a} \extchoice \bar{a}_{S}) \underset{a_{S}\#2}{\parallel} (a \extchoice a_{S}) 
\end{align*}
Thanks to ~$\hideCSP\set{\tau[a,\bar{a}]}$ being vacuous, there will be two possible synchronisations on $a_{S}$, corresponding to the original CCS behaviour.	
\end{example}
\par 
The following abbreviation translates CCSTau into CSPmn.
\begin{definition}[CCSTau to CSPmn]\label{def:ccstau2cspmn}
Let $P$ be a CCSTau process. Then: 
\begin{align*}
t2csp_{3}(P) ~&\defs~ (tl_{3} \circ conm \circ g^{2}(P))\hideCSP\set{tau}
\end{align*}  
\end{definition}
\par 
%
\paragraph{Link CCS-to-CSPmn.}
We obtain the translation from CCS to CSP by translating CCS into CCSTau first, using $c2ccs\tau$ (Def.\ref{c2ccstau-par-def}), then translating CCSTau into CSPmn, using $t2csp_{3}$ (Def.\ref{def:ccstau2cspmn}), and finally hiding every $a_{S}$ synchronisation event.
\begin{definition}[CCS to CSPmn]\label{def:ccs2csp:m/n}
Let $P$ denote a CCS process. Then: 
$$ ccs2csp_{3}(P) ~\defs~ (t2csp_{3} \circ c2ccs\tau(P))\hideCSP\set{a_{S} | a_{S} \in \mathcal{A}(t2csp_{3} \circ c2ccs\tau(P))} $$
\end{definition}
%
\begin{example}\label{eg:ccs2cspmn-par}
The translation of CCS binary synchronisation into CSPmn can be illustrated succinctly as follows:
\begin{align*}
& ccs2csp_{3}(a.0 | \bar{a}.0) \tag{ccs2csp3-def.\ref{def:ccs2csp:m/n}}\\
=&~ \big(t2csp_{3} \circ c2ccs\tau(a.0 | \bar{a}.0) \big)\hideCSP\set{a_{S} |..} \tag{\ref{c2ccstau-par-def}}\\
=&~ t2csp_{3}\big((a.0 |_{_{T}} \bar{a}.0)\hideT\{\tau[a|\bar{a}]\}\big)\hideCSP\set{a_{S}} \tag{t2csp3-def.\ref{def:ccstau2cspmn}}\\
=&~ tl_{3} \circ conm \circ g^{2}(\nullset, (a.0 |_{_{T}} \bar{a}.0)\hideT\{\tau[a|\bar{a}]\})\hideCSP\set{tau}\hideCSP\set{a_{S}} \tag{g2-def.\ref{def:g(alpha):m/n}}\\
=&~ tl_{3} \circ conm \Big(\big((a_{}.0 + a_{S}.0) |_{_{T}} (\bar{a}_{}.0 + \bar{a}_{S}.0)\big)\hideT\set{\tau[a|\bar{a}]}\Big)\hideCSP\set{tau}\hideCSP\set{a_{S}} \tag{conm-def.\ref{def:conm}}\\
=&~ tl_{3}\Big(\big((a_{}.0 + a_{S}.0) |_{_{T}} (\bar{a}_{}.0 + a_{S}.0)\big)\hideT\set{\tau[a|\bar{a}]}\Big)\hideCSP\set{tau}\hideCSP\set{a_{S}} \tag{tl3-def.\ref{def:tl3}, CSP}\\
=&~ \big((a_{} \extchoice a_{S} \leadsto STOP) \underset{\{a_{S}\#2\}}{\parallel} (\bar{a}_{} \extchoice a_{S} \leadsto STOP)\big)\hideCSP\set{tau, a_{S}}   
\end{align*}
\end{example}
\begin{example}
The translation of recursion with nested parallel can be illustrated as follows. \linebreak 
Let $P \defs \rec{X}(a | \bar{a}.X)$ (or equiv. $P \defs a.0 ~|~ \bar{a}.P$) be a CCS process. Then,
$ ix(P) = a_{1} ~|~ a_{2}.ix_{\set{3..}}(P) $, 
where $ix_{\set{3..}}$ denotes that indexing excludes indices $1$ and $2$. Let us unfold $P$ one step, then:
\begin{align*}
P &= a ~|~ \bar{a}.(a ~|~ \bar{a}.P) \\
 ix(P) &= a_{1} ~|~ \bar{a}_{2}.(a_{3} ~|~ \bar{a}_{4}.ix_{\set{5..}}(P)) 
\end{align*} 
The synchronisation pairs are thus $(a_{1}, \bar{a}_{2}), (a_{1}, \bar{a}_{4}), ..$, that is, the set $\set{(a_{1}, \bar{a}_{2k}) | k \geq 1}$.
Then:
$$
g^{\ast}(P) = (a_{1} + \underset{k \geq 1}{\Sigma} a_{1*2k}) ~|~ (\bar{a}_{2}+\bar{a}_{12}).g^{\ast}(P)
$$
We will not be able to generate all the $a_{1*2k}$ indices since recursion is unbounded. For closure, we give the \emph{temptative} translation of $P$ with $ccs2csp$: 
	\footnote{We are lucky that we can tell in advance what the synchronisation indices are, because process $P$ is a simple case.}
$$
ccs2csp(P) = \big( (a \extchoice \underset{k \geq 1}{\Extchoice} a_{1*2k}) \underset{\set{a_{1*2k} | k \geq 1}}{\parallel} (\bar{a}_{2} \extchoice a_{12}) \leadsto ai2a \circ t2csp \circ c2ccs\tau(P) \big)\hideCSP\set{a_{ij} | ..}
$$
In contrast, let us define $ccs2csp_{3}(P)$. Then:
\begin{align*}
g^{2}(P) &= (a + a_{S}) ~|~ (\bar{a} + \bar{a}_{S}).g^{2}(P) \\
&= (a + a_{S}) ~|~ (\bar{a} + \bar{a}_{S}).\big((a + a_{S}) ~|~ (\bar{a} + \bar{a}_{S}).g^{2}(P) \big)
\end{align*}
We can unfold $P$ multiple times, we only ever generate a single name for synchronisation. Then:
$$
ccs2csp_{3}(P) = \big((a \extchoice a_{S}) \underset{a_{S}\#2}{\parallel} (\bar{a} \extchoice a_{S}) \leadsto t2csp_{3} \circ c2ccs\tau(P) \big)\hideCSP\set{a_{S}}
$$
\end{example}

%% file: Gstar2MN.tex
\section{Gstar Implements 2/n-Synchronisation}
We discuss here the relation between $g^{\ast}$-renaming (\cite{BEK21}) and m-among-n synchronisation (\S\ref{sec:cspmn}) approaches. \\
\par 
Recall, function $g^{\ast}$ (Def.\ref{def:ccs2csp}, \cite{BEK21}) computes for a CCSTau process $P$ all the substitute names corresponding to distinct synchronisation possibilities of $P$ with its environment, plus interleaving. We have proposed an alternative solution based on extending CSP with 2-among-n synchronisation, derived from first extending CSP with m-among-n synchronisation. Whilst this second solution is more elegant than the gstar-renaming one, the problem of its \emph{immediate} implementability in a tool like FDR has been raised. 
\par 
Given the current version of FDR, m-among-n synchronisation cannot be implemented directly. We remark, however, that one effect of m-among-n synchronisation is to select, using non-deterministic choice, the $m$ processes that are allowed to synchronise; effect which is precisely what function $g^{\ast}$ achieves through renaming. We discuss how to relate both results. \\%
\par
Let us refer by CSPgstar the CSP process expressions resulting from translation $ccs2csp$. We can translate CSPgstar expressions into CSPmn expressions as follows.
\begin{definition}[gstar2m/n]\label{def:gstar-to-mn}
Let $a_{ij}$ be an $g^{\ast}$ name, $a_{S}$ an $g^{2}$ name. Then:
$
g^{\ast}2g^{2} \defs \set{\tau \mapsto \tau, a_{ij} \mapsto a_{S} }
$ 
\end{definition}
While $g^{\ast}2g^{2}$ is a simple renaming function, its application to CSP processes is modified specifically for the parallel operator such as to map $\underset{\set{a_{ij}}}{\parallel}$ unto $\underset{\set{a_{S}\#2}}{\parallel}$ (instead of $\underset{\set{a_{S}}}{\parallel}$).
\begin{definition}
Let $P$ be a CSP process.
\begin{align*}
\begin{aligned} 
g^{\ast}2g^{2}(STOP) &\defs STOP \\[+2ex]
g^{\ast}2g^{2}(\alpha \leadsto P) &\defs g^{\ast}2g^{2}(\alpha) \leadsto g^{\ast}2g^{2}(P) \\
g^{\ast}2g^{2}(P \intchoice Q) &\defs g^{\ast}2g^{2}(P) \intchoice g^{\ast}2g^{2}(Q) 
\end{aligned}
&\qquad \begin{aligned} 
g^{\ast}2g^{2}(P \underset{\set{a_{ij}}}{\parallel} Q) &\defs g^{\ast}2g^{2}(P) \underset{\set{a_{S}\#2}}{\parallel} g^{\ast}2g^{2}(Q) \\
g^{\ast}2g^{2}(P\hideCSP{B}) &\defs g^{\ast}2g^{2}(P)\hideCSP{g^{\ast}2g^{2}(B)} \\
g^{\ast}2g^{2}(P \extchoice Q) &\defs g^{\ast}2g^{2}(P) \extchoice g^{\ast}2g^{2}(Q) 
\end{aligned}
\end{align*}
\end{definition}
\begin{theorm}
Let $P$ be a CCS processes. Then: $g^{\ast}2g^{2} \circ ccs2csp(P) = ccs2csp_{3}(P)$.
\begin{proof}
By induction on the structure of CCS processes. When $P$ does not mention CCS parallel, the proof is straightforward. We develop the proof for the parallel case only. We have:
\begin{align*}
&g^{\ast}2g^{2} \circ ccs2csp(P ~|~ Q) \tag{ccs2csp-def.\ref{def:ccs2csp}} \\
=~&
g^{\ast}2g^{2} \circ ai2a \circ (t2csp(P) \underset{\set{a_{ij}}}{\parallel} t2csp(Q))\hideCSP\set{tau, a_{ij}} \tag{CSP hide law}  \\
=~&
g^{\ast}2g^{2} \circ (ai2a \circ t2csp(P)\hideCSP\set{tau} \underset{\set{a_{ij}}}{\parallel} ai2a \circ t2csp(Q)\hideCSP\set{tau}) \hideCSP\set{a_{ij}} \tag{g*2g2-def.\ref{def:gstar-to-mn}} \\
=~&
\big(g^{\ast}2g^{2} \circ ai2a \circ t2csp(P)\hideCSP\set{tau} \underset{\set{a_{S}\#2}}{\parallel} g^{\ast}2g^{2} \circ ai2a \circ t2csp(P)\hideCSP\set{tau}\big)\hideCSP\set{a_{S}} \tag{Induction Hyp., ccs2csp3-def.\ref{def:ccs2csp:m/n}} \\
=~&ccs2csp_{3}(P ~|~ Q)
\end{align*}
\end{proof}
\end{theorm}
We say that $g^{\ast}$ implements 2-among-n synchronisation.

%% file: Conclusion.tex
\section{Conclusion and Future Work}\label{conclusion}
\cite{BEK21} proposes a translation of CCS into CSP based on the $g^{\ast}$-renaming approach whereby if two processes can synchronise on an action $b$, then a name unique to these two processes, say $b_{ij}$, is generated to substitute $b$. Thus, if more than two processes could initially synchronise on $b$, only two processes will ever be able to synchronise on $b_{ij}$ after application of $g^{\ast}$.
\par
In this paper, we propose an alternative, the m-among-n synchronisation approach, whereby we first extend CSP multiway synchronisation (or n-among-n) to m-among-n synchronisation (extension called CSPmn), from which we derive 2-among-n or binary synchronisation for CSP processes. We then translate CCS binary synchronisation into CSPmn binary synchronisation. Unlike the $g^{\ast}$-renaming approach, the m/n-approach is not limited by parallel under recursion since we can generate a single synchronisation name, say $a_{S}$, independently of the number of processes meant to synchronise on $a_{S}$.
\par 
We have also shown that CSPmn is a conservative extension of CSP (viz., preserves CSP axioms) by defining (CSPmn) m-among-n synchronisation in terms of both (CSP) multiway (or n-among-n) synchronisation and relational renaming.
\par
We are tempted to affirm that m-among-n synchronisation is more expressive than both 2-among-n and n-among-n synchronisation. However, Hatzel et al. \cite{Hat+15} propose an encoding from CSP into CCS whereby they encode CSP multiway synchronisation based on CCS binary synchronisation. Our work suggests that in trying to translate CSP into CCS, it would be easier to extend CCS with multiway synchronisation, as we have done here for CSP. Other works on the translation from CSP into CCS include \cite{Ast+81}, \cite{Broo83}, \cite{Hen+81}, and \cite{Glab12}.
\par 
We have proposed here the translation from CCS to CSP only. The main reason for this is our interest in using CSP tools such as FDR for reasoning about CCS processes. With regard to this concern, the $g^{\ast}$-renaming approach is more readily implementable than the m/n-approach. The latter would require extending FDR with semantics (viz.\ rules) for m-among-n synchronisation. Alternatively, m-among-n synchronisation can be implemented using function $g^{\ast}_{\#}$ (Def.\ref{def:gs-sharp:proc}), however, with the limitation on parallel under recursion similar to $g^{\ast}$ (cf.\ \cite{BEK21}). Mechanising our results in Isabelle theorem prover is also to be explored in the future.

%% file: App_ProofCSPmnConserve.tex
\section{Proof that CSPmn is a Conservative Extension}\label{app:proof-cspmn-corr}
In order to prove that CSPmn is conservative, we need to define some auxillary functions. 
First, we uniquely index the prefixes of CSP processes.
\begin{propty}\label{pty:ix-csp}
Let $P$ be a CSP process.
    \begin{align*}
    &\begin{aligned}
    ix(STOP) &= STOP \\
    ix(a \leadsto P) &= a_{i} \leadsto ix_{-i}(P) \\
    ix(P \intchoice Q) &= ix_{1}(P) \intchoice ix_{2}(Q) \\
    ix(P \extchoice Q) &= ix_{1}(P) \extchoice ix_{2}(Q) 
    \end{aligned}
    &\begin{aligned}
    ix(P \underset{a\#m}{\parallel} Q) &= ix_{1}(P) \underset{B}{\parallel} ix_{2}(Q) \\
    B &\defs \set{a_{i}\#m | a_{i} \in \mathcal{A}(ix_{1}(P)) \cup \mathcal{A}(ix_{2}(Q))} \\
    ix(P\hideCSP\{a\}) &= ix(P)\hideCSP\{a_{i} | a_{i} \in \mathcal{A}(ix(P))\} \\
    ix(\rec{X}P) &= \rec{X}ix(P) \\
    ix(X) &= X
    \end{aligned}
    \end{align*}
      where $ix_{-i}$ is some indexing scheme which does not assign the $i$-index,
      and $ix_1, ix_2$ are indexing schemes that assign disjoint indices.
    \end{propty}

Then, using $ix$-generated indices we generate unique synchronisation indices. Given a set $\set{a_{i}}$ of parallel prefixes and a number $m$ of processes meant to synchronise together, $g^{\ast}_{a_{i}\#m}$ generates a unique synchronisation index $a_{i_{1}..i_{m}}$. 

\begin{definition}\label{def:gsharp:alpha}
    Let $S, B$ denote sets of indexed events.
\begin{align*}
&g^{\ast}_{a_{i_{1}}\#m}(S, a_{i_{1}}) \defs 
    \set{ a_{i_{1}..i_{m}} ~|~ i_{1} < .. < i_{m}, \set{a_{i_{k}} ~|~ 1 < k \leq m} \subseteq S } ~\cup 
    \set{ a_{i_{m}..i_{1}} ~|~  i_{m} < .. < i_{1}, \set{a_{i_{k}} ~|~ 1 < k \leq m} \subseteq S } \\
&g^{\ast}_{\set{a_{k}\#m_{k} | k \in \nats}}(S, a_{i}) \defs 
\begin{cases}
    a_{i} & a_{i} \notin \set{a_{k} ~|~ k \in \nats} \\
    g^{\ast}_{a_{i}\#m_{i}}(S, a_{i}) & ~
\end{cases}
\end{align*}
\end{definition}

Although $g^{\ast}_{a\#m}$ denotes relational renaming, we overload its application to processes such that it translates $\underset{a\#m}{\parallel}$ into $\underset{a}{\parallel}$. This corresponds to the following.

\begin{definition}\label{def:gs-sharp:proc}
Let $P$ be an $ix$-indexed CSP processes. Let $S$ be a set of $ix$-indexed events. 
Let $a\#m$ denote the set $\set{a_{k}\#m_{k} ~|~ k \in \nats}$, $b\#n$ the set $\set{b_{j}\#n_{j} ~|~ j \in \nats}$. 
Let $g^{\ast}_{a\#m}(P) \defs g^{\ast}_{a\#m}(\nullset, P) $.
\begin{align*}
g^{\ast}_{a\#m}(S, STOP) &\defs STOP 
\hspace{5.6cm}
g^{\ast}_{a\#m}(S, a \leadsto P) \defs \underset{b \in g^{\ast}_{a\#m}(a)}{\Sigma}b \leadsto g^{\ast}_{a\#m}(S, P) 
\\
g^{\ast}_{a\#m}(S, P \intchoice Q) &\defs g^{\ast}_{a\#m}(S, P) \intchoice g^{\ast}_{a\#m}(S, Q) 
\hspace{2.9cm}
g^{\ast}_{a\#m}(S, P \extchoice Q) \defs g^{\ast}_{a\#m}(S, P) \extchoice g^{\ast}_{a\#m}(S, Q) 
\\
g^{\ast}_{a\#m}(S, \rec{X}P) &\defs \rec{X}g^{\ast}_{a\#m}(S, P) 
\hspace{4.95cm}
g^{\ast}_{a\#m}(S, X) \defs X 
\\
g^{\ast}_{a\#m}(S, P\hideCSP\{a\}) &\defs g^{\ast}_{a\#m}(S, P)\hideCSP{g^{\ast}_{a\#m}(S, a)} 
\\
g^{\ast}_{a\#m}(S, P \underset{b\#n}{\parallel} Q) &\defs g^{\ast}_{a\#m ~\cup~ b\#n}(S \cup \mathcal{A}(Q), P) \underset{B}{\parallel} g^{\ast}_{a\#m ~\cup~ b\#n}(S \cup \mathcal{A}(P), Q) 
\\
B &\defs \bigcup \set{ g^{\ast}_{a\#m ~\cup~ b\#n}(S \cup \mathcal{A}(Q), b_{j}) | b_{j} \in \mathcal{A}(P)} ~\cup 
        \bigcup \set{ g^{\ast}_{a\#m ~\cup~ b\#n}(S \cup \mathcal{A}(P), b_{j}) | b_{j} \in \mathcal{A}(Q)} 
\end{align*}
\end{definition}

When $a\#m$ denotes the empty set, we write $g^{\ast}_{\#}$ for the corresponding function $g^{\ast}_{a\#m}$. Then, the translation of CSPmn into CSP is given by the following.

\begin{definition}\label{def:mn2csp}
Let $P$ be a CSPmn process. 
$mn2csp(P) \defs g^{\ast}_{\#} \circ ix(P)$
\end{definition}

The following theorem establishes a labelled operational correspondence (Def.~\ref{def:opCorrespond}), which turns out a strong bisimulation (Def.~\ref{def:strongbisim}), between CSPmn and CSP. 

\begin{theorm}\label{thrm:cspmn-correct}
Let $P$ be a CSPmn process. Let $I$ denote a given sequence of natural numbers.
\begin{enumerate}
\item If $P \xrightarrow{a} P'$ then $\exists I: mn2csp(P) \xrightarrow{a_{_{I}}} Q$ and $Q \equiv mn2csp(P')$
\item If $mn2csp(P) \xrightarrow{a_{_{I}}} Q$ then $\exists!P': P \xrightarrow{a} P'$ and $Q \equiv mn2csp(P')$
\end{enumerate}
\begin{proof}
When $P$ does not mention $\underset{a\#m}{\parallel}$, $mn2csp$ behaves like the identity function, hence the theorem holds. By induction, we prove the case for parallel. \\
~\\
(Thrm.\ref{thrm:cspmn-correct}.1.)
[Induction step:Parallel]. Let $P_{1} \xrightarrow{a} P'_{1}$. Let $P_{2},..,P_{n}$ denote processes such that $m-1$ among them can perform an $a$-transition. 
For ease, we select one such combinations, $P_{2} .. P_{m}$. 
The following result applies for all possible combinations. 
---(Hyp-combine)--- 
Then, by M/N-IndxIfacePar rule (\S\ref{sec:cspmn}), 
\begin{align*}
P_{1} \underset{a\#m}{\parallel} .. \underset{a\#m}{\parallel} P_{n} 
    ~~\xrightarrow{a}~~ 
P'_{1} \underset{a\#m}{\parallel} P'_{2} ..  \underset{a\#m}{\parallel} P'_{m} \underset{a\#m}{\parallel} P_{m+1} \underset{a\#m}{\parallel} .. \underset{a\#m}{\parallel} P_{n}
\end{align*}
Assume for each $P_{i}$ that every occurrence of $a$ in $P_{i}$ is indexed into $a_{i}$. (The following applies even if we separate $i$ into distinct indices, e.g., $i_{1}, i_{2}, ..$, as many as there are of instances of $a$ in $P_{i}$.) 
---(Hyp-indx)---
Then, by (Hyp-combine), (Hyp-indx), and Def.\ref{def:gs-sharp:proc}, $g^{\ast}_{\#}(a) = a_{12..m}$ and:
\begin{align*}
mn2csp(P_{1} \underset{a\#m}{\parallel} .. \underset{a\#m}{\parallel} P_{n}) = P_{1}[a_{12..m}/a] \underset{a_{12..m}}{\parallel} .. \underset{a_{12..m}}{\parallel} P_{m}[a_{12..m}/a] \underset{\nullset}{\parallel} P_{m+1} \underset{\nullset}{\parallel} .. \underset{\nullset}{\parallel} P_{n}
\end{align*}
By IndxIfacePar rule (\S\ref{sec:cspmn}) and definition of renaming (Tab.\ref{tab:CSPrules}):
\begin{align*}
P_{1}[a_{12..m}/a] \underset{a_{12..m}}{\parallel} .. \underset{a_{12..m}}{\parallel} P_{m}[a_{12..m}/a] &\underset{\nullset}{\parallel} P_{m+1} \underset{\nullset}{\parallel} .. \underset{\nullset}{\parallel} P_{n} 
\quad\xrightarrow{a_{12..m}} \\ 
&P'_{1}[a_{12..m}/a] \underset{a_{12..m}}{\parallel} ..  \underset{a_{12..m}}{\parallel} P'_{m}[a_{12..m}/a] \underset{\nullset}{\parallel} P_{m+1} \underset{\nullset}{\parallel} .. \underset{\nullset}{\parallel} P_{n} 
\end{align*}
Then, by induction hypothesis. \\
~\\
(Thrm.\ref{thrm:cspmn-correct}.2.) 
[Induction step: Prallel.] Let $mn2csp(P) \xrightarrow{a_{_{I}}} Q$. 
By Par rule, 
$
mn2csp(P) \underset{a_{_{I}}}{\parallel} mn2csp(P_{2}) \xrightarrow{a_{_{I}}} Q \underset{a_{_{I}}}{\parallel} mn2csp(P_{2})$, 
$a_{_{I}} \notin \mathcal{A}(mn2csp(P_{2}))
$. 
By induction hypothesis, 
$\exists! P': P \xrightarrow{a} P'$ and $Q \equiv mn2csp(P')$. 
Then, by Par rule, $P \underset{a\#m}{\parallel} P_{2} \xrightarrow{a} P' \underset{a\#m}{\parallel} P_{2}$. 
Moreover, $Q \underset{a_{_{I}}}{\parallel} mn2csp(P_{2}) \equiv mn2csp(P') \underset{a_{_{I}}}{\parallel} mn2csp(P_{2}) = mn2csp(P' \underset{a\#m}{\parallel} P_{2})$, by Def.\ref{def:mn2csp}.
\end{proof}
\end{theorm}

As a consequence, when m-among-n CSPmn processes, $\underset{a\#m, j}{\Parallel}P_{j}$, will synchronise on $a$, m-among-n CSP processes, $\underset{a_{12..m}, j}{\Parallel} P_{j}[a_{12..m}/a]$, will synchronise on $a_{12..m}$, where $12..m$ denotes any combination of $m$ potential synchronising processes. We say that $mn2csp$ implements m-among-n synchronisation.